\documentclass{jkas}

\def\year{2026} % publication year
\def\volume{59} % journal volume
\def\issue{1} % journal issue
\def\beginpage{33} % first page of article
\date{Received October 9, 2025; Accepted January 28, 2026; Published February 10, 2026}

\title{%
A redshift survey of the nearby galaxy cluster Abell 2199 : No upturn of the faint-end slope of galaxy luminosity function
}

\author[1]{Jong-In Park}{0009-0009-4334-5598}
\author[2]{Hyunmi Song}{0000-0002-4362-4070}
\author[1,3]{Ho Seong Hwang}{0000-0003-3428-7612}
\affil[1]{Astronomy Program, Department of Physics and Astronomy, Seoul National University, 1 Gwanak-ro, Gwanak-gu, Seoul 08826, Republic of Korea}
\affil[2]{Department of Astronomy and Space Science, Chungnam National University, Daejeon 34134, Republic of Korea}
\affil[3]{SNU Astronomy Research Center, Seoul National University, 1 Gwanak-ro, Gwanak-gu, Seoul 08826, Republic of Korea}

\def\corrauthor{%
H. Song, \email{hmsong@cnu.ac.kr} and H. S. Hwang, \email{hhwang@astro.snu.ac.kr}
}

\def\runningauthor{%
Jong-In Park
}

\def\runningtitle{%
Luminosity Function of Abell 2199 Cluster Galaxies
}

\def\keywords{%
galaxies: clusters: individual (Abell 2199) --- galaxies: luminosity function
}

\def\abstracttext{%
We determine the galaxy luminosity function of cluster galaxies in the nearby galaxy cluster Abell 2199 (A2199), focusing on the faint-end slope down to $M_r \sim -14.5$. To achieve this, we augment the existing dataset by adding redshift data from our deep MMT/Hectospec survey and from the Dark Energy Spectroscopic Instrument (DESI), significantly improving the spectroscopic completeness down to $r_{\mathrm{petro},0} = 20.8$ within the central $30^\prime$ region. The resulting luminosity function is well described by a Schechter function with a characteristic magnitude $M^* = -21.30\pm{0.27}$ and a faint-end slope $\alpha = -1.23\pm{0.05}$. This faint-end slope is consistent with those measured in the nearby Coma and Virgo clusters and in a cluster from the TNG50 cosmological simulation, and is slightly shallower than that of field galaxies. These findings indicate that the previously claimed steep faint‐end upturn (with $\alpha \sim -2$) in nearby galaxy clusters is not supported. Instead, they indicate that environmental processes in dense cluster cores does not seem to trigger the formation or survival of low-mass galaxies, thereby preventing a steep faint-end upturn in the luminosity function.
}

\begin{document}
\jkashead %% set title, authors, abstract, etc.

%%%%%%%%%%%%%%%%%%%%%%%%%%%%%%%%%%%%%%%%%%%%%%%%%%%%%%%%%%%%%%%%%%%%%
%%% BEGIN MAIN TEXT HERE %%%%%%%%%%%%%%%%%%%%%%%%%%%%%%%%%%%%%%%%%%%%
%%%%%%%%%%%%%%%%%%%%%%%%%%%%%%%%%%%%%%%%%%%%%%%%%%%%%%%%%%%%%%%%%%%%%

\section{Introduction} \label{sec:intro}

Galaxy clusters are the largest gravitationally bound structures in the universe, composed of multiple components with different mass fractions: approximately 80–95\% dark matter, 5–20\% intracluster medium (ICM), and 0.5–3\% in galaxies \citep{Lin2003, Ettori2009}.

The galaxy luminosity function serves as a foundational tool for probing galaxy formation and evolution \citep{McNaughtRoberts2014, Moretti2015}. The luminosity function traces the mass function of dark matter halos where galaxies reside. Yet, discrepancies exist between the observed luminosity function and the predicted dark matter halo mass function, particularly at the bright and faint ends, where galaxies are notably underrepresented \citep{Driver2009, Kim2023a}. These mismatches can be attributed to various astrophysical mechanisms, such as gas cooling, cosmic reionization, feedback processes, mergers, and thermal conduction \citep{Benson2003,Cooray2005,Croton2006,Park2018}. 

The luminosity functions of field galaxies (field luminosity function from now on) are known to have relatively shallow faint-end slopes (i.e., $\alpha > -1.3$): as shown by the Sloan Digital Sky Survey (SDSS, \citealp{Blanton2005}), the Cosmic Evolution Survey (COSMOS, \citealp{Liu2008}), the UKIRT Infrared Deep Sky Survey (UKIDSS, \citealp{Lawrence2007, Smith2009}), and the Smithsonian Hectospec Lensing Survey (SHELS, \citealp{Geller2012}). In contrast, the luminosity functions in clusters based only on photometry shows significantly steeper slopes \citep{ Driver1994,DePropris1995,Banados2010,Moretti2015}. For instance, \citet{Popesso2006} reported that a steep upturn with $\alpha = -2.19$ ($r$-band) from a composite luminosity function of 69 clusters, suggesting that physical mechanisms like tidal shielding or UV protection could favor the survival of dwarf galaxies in cluster environments. Conversely, not all luminosity functions based on photometry exhibit steep faint-end slopes. \citet{Andreon2002} and \citet{Andreon2006} constructed luminosity functions for the Coma cluster and A1185, and found relatively shallow faint-end slopes. Furthermore, \citet{Andreon2005} reported significant cluster-to-cluster variations in the faint-end slope, for example between AC 114 and AC 118.

However, photometry-based luminosity function studies are susceptible to background contamination, particularly at faint magnitudes. \citet{Valotto2001} showed that traditional background subtraction methods do not completely eliminate background sources using mock galaxy simulations when clusters are embedded in larger structures. Spectroscopic surveys offer more reliable constraints by directly measuring redshifts. 

To our knowledge, the Virgo cluster remains the only system for which sufficiently deep spectroscopic data exist ($M^* + 8$), with both spectroscopic and photometric results supporting a shallow faint-end slope \citep{Rines2008, Ferrarese2016}. To draw a general conclusion on the faint-end slope of cluster luminosity function, it is necessary to increase the sample size with a particular focus on spectroscopy to avoid the contamination.

Abell 2199 (A2199), a nearby ($z = 0.03$), rich, and X-ray-bright cluster, is an ideal target for such an investigation. A2199 has $M_{200} \sim 10^{14.53}\ \mathrm{M}_\odot$, $R_{200} \sim 1.42\ \mathrm{Mpc}$ (HeCS-omnibus catalog, \citealp{Sohn2020}). A2199 contains a central cD galaxy (NGC 6166) which shows radio jet activity \citep{Nulsen2013}, and forms a supercluster with neighboring infalling groups \citep{Rines2001}. Its proximity, structural complexity, and previous extensive observations—including SDSS \citep{York2000}, WISE \citep{Wright2010}, and X-ray data from ROSAT and Suzaku (\citealp{Voges1999}; Tamura, in prep.)—make it a powerful laboratory for studying the interplay between galaxies, the ICM, and dark matter \citep[e.g.][]{Rines2001, Rines2002, Hwang2012, Lee2015}.

In this regard, \citet{Song2017} present the results from the first MMT/Hectospec redshift survey in the central $R < 30^\prime$ region of A2199. They increase the magnitude limit of the spectroscopic completeness to $m_{r,\mathrm{Petro},0} = 20.2$ (where the completeness drops below 50\%). They construct the luminosity function for galaxies with $M_r < -15$ and report the best-fit Schechter parameters of $M^* = -21.68 \pm 0.80$ and $\alpha = -1.26 \pm 0.06$. They find no evidence of a faint-end upturn in the luminosity function of A2199 at $M_r < -15$. 

The absence of a faint-end upturn in cluster luminosity function may result from a redshift survey that is not deep enough to detect the upturn that starts to appear $M_r\sim -16$ in \citet{Popesso2006}. We therefore perform second MMT/Hectospec redshift survey of A2199 and push the magnitude limit by 0.6 mag (i.e., $m_{r,\mathrm{Petro},0}=20.8$, $M_{r,\mathrm{Petro},0} \sim -14.5$)

Here, we present the results from this new redshift survey along with Data Release 1 of the Dark Energy Spectroscopic Instrument (DESI DR1; \citealp{AbdulKarim2025}). We construct a luminosity function of A2199 down to $M_r \sim -14.5$.  We focus on the measurement of the faint-end slope of the luminosity function, and compare with those of other local clusters, field, and a simulated cluster. 
 
This paper is organized as follows. In Section~\ref{sec:data}, we describe our photometric and spectroscopic data used for constructing luminosity functions, as well as the membership determination process. Section~\ref{sec:luminosity_function_a2199} describes the luminosity function of A2199. In Section~\ref{sec:discussion}, we introduce luminosity functions from several comparison samples, including local massive clusters (Coma and Virgo), a local field sample, and the cluster in the TNG50 simulation. We then compare these luminosity functions and discuss the faint-end slope in local clusters. Finally, in Section~\ref{sec:conclusion}, we summarize our conclusions. Throughout this paper, we adopt a flat Lambda cold dark matter ($\Lambda$CDM) cosmological parameters with $H_0 = 70\mathrm{km\,s^{-1}\,Mpc^{-1}}$, $\Omega_\Lambda = 0.7$, and $\Omega_m = 0.3$.

\section{Data} \label{sec:data}

\subsection{Photometric Data} \label{sec:photdat}
\begin{table*}[t]
  \centering
  \caption{Summary of MMT/Hectospec observation fields.}
  \label{tab:MMT_fields}
  \begin{tabular}{ccccccc}
    \toprule
    Field ID & R.A.\ $(^\circ)$ & Decl.\ $(^\circ)$ & Date & Exposure (min) & Number of Redshifts & Number of Targets \\
    \midrule
    % Row 1
    {a2199a19\_1}  & {247.1582} & {39.5487} & {2019 April 29} & {60} & {181} & {264} \\
    % Row 2
    {a2199a19\_2}  & {247.1582} & {39.5487} & {2019 March 9} & {75} & {149} & {262} \\
    % Row 3
    {a2199a19\_3}  & {247.1582} & {39.5487} & {2019 April 28} & {75} & {213} & {262} \\
    % Row 4
    {a2199a19\_4}  & {247.1582} & {39.5487} & {2019 May 1} & {105} & {238} & {264} \\
    \bottomrule
  \end{tabular}
\end{table*}

\begin{figure*}[t]
    \centering    \includegraphics[width=
    \textwidth]{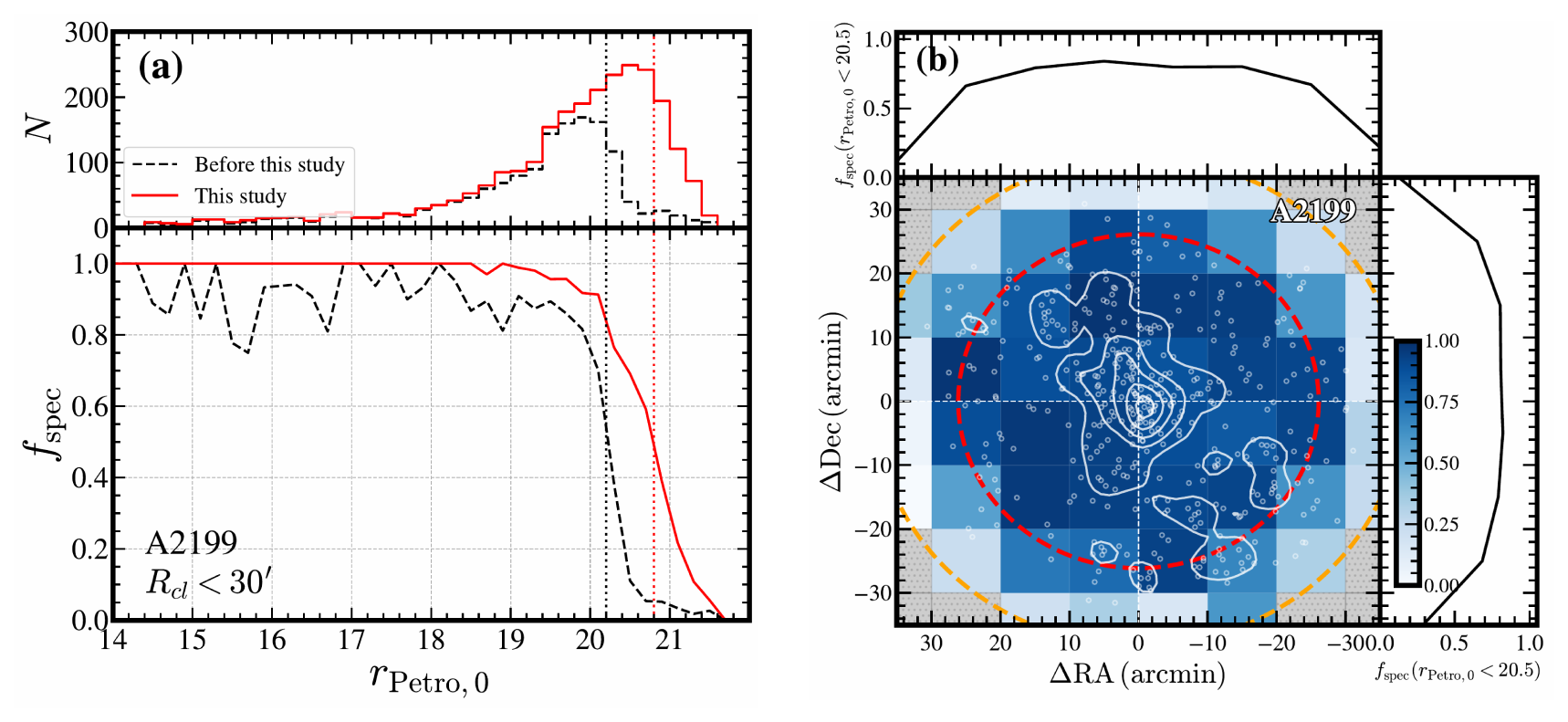}
        \caption{(a)
        (Top) Histogram of galaxies within 30$^\prime$ from the cluster center for all galaxies with spectroscopy, shown as a function of $r$-band Petrosian magnitude. The black dashed and red solid lines show the results before and after this study, respectively.
        (Bottom) Differential spectroscopic completeness, $f_\mathrm{spec}$, as a function of magnitude. The black and red dotted lines indicate the magnitude limits where the spectroscopic completeness falls below 50\%, before and after this study, respectively.
        (b) Spatial spectroscopic completeness of galaxies with \( m_{r,\mathrm{Petro},0} < 20.5 \) in the A2199 field. Marginal spectroscopic completeness along the spatial axes—right ascension (top) and declination (right)—is shown in the subpanels. Darker colors indicate higher spatial spectroscopic completeness. White circles show the member galaxies, whose density distribution is indicated by white contours. Red and orange circles indicate the \( R_{500} \) and \( R_{200} \) of A2199, respectively.}
    \label{fig:fspec_a2199}
\end{figure*}

We use the photometric data from the SDSS Data Release (DR) 17 \citep{Abdurrouf2022} as our base photometric catalog. The SDSS catalog provides right ascension, declination, as well as fiber, Petrosian, and model magnitudes in $ugriz$ bands. We use the Petrosian magnitudes, which are widely used for galaxy photometry. We apply foreground extinction corrections to all magnitudes using the extinction values provided in the SDSS catalog (\texttt{extinction\_g}, \texttt{extinction\_r}, etc.). These values are derived from the dust maps of \citet{Schlegel1998} using the conversion coefficients defined in the SDSS Early Data Release \citep{Stoughton2002}. Among galaxies without spectroscopic redshifts, brighter ones are assigned higher priority in follow-up spectroscopic observations.

To enable direct comparisons between galaxies at varying redshifts, we apply the K-correction \citep{Oke1968} when computing the absolute magnitudes of the galaxies with spectroscopic redshifts. This correction transforms the photometric data from the observed frame to the common reference frame, thereby facilitating meaningful comparisons of galaxy properties such as luminosity and color across different redshifts. We use the \texttt{kcorrect} package developed by \citet{Blanton2007}. We convert the SDSS Petrosian \( ugriz \) magnitudes and their associated errors into linear flux units (maggies) and corresponding inverse variances. We set the \texttt{nredshift} parameter to 6000. We apply a \texttt{band\_shift} of 0.1, which effectively shifts the filter response curves to correspond to a rest-frame redshift of \( z = 0.1 \). We adopt the cosmological model described in Section~\ref{sec:intro}.

\subsection{Spectroscopic Data} \label{sec:spec_data}

To construct a deeper spectroscopic data for the central region of A2199 we perform an additional redshift survey with Hectospec. We first collect the redshift data from our own redshift survey with Hectospec on the MMT 6.5m telescope \citep{Fabricant1998, Fabricant2005}. Hectospec is a 300-fiber multi-object spectrograph with a circular field of view (FOV) of $1^\circ$ in diameter. We use the 270 line mm$^{-1}$ grating, which provides a dispersion of 1.2~\AA~pixel$^{-1}$ and a resolution of $\sim$6~\AA. We observe four fields, each with three exposures of 20 minutes, and obtain spectra covering the wavelength range 3500--9150~\AA. All fields are centered on the X-ray center of A2199 (R.A. = $247^\circ.1582$, decl. = $39^\circ.5487$; \citealt{Boehringer2000}). Table \ref{tab:MMT_fields} summarizes the information about the observed MMT/Hectospec fields. The targets are selected from the photometric data of SDSS DR17 with $m_{r,\mathrm{Petro},0}<21$, regardless of colors, giving higher priority to brighter galaxies lacking spectroscopic redshifts. 

We reduce the Hectospec spectra using HSRED v2.0, an updated version of the reduction pipeline originally developed by Richard Cool. We determine redshifts with the RVSAO package \citep{Kurtz1998} by cross-correlating the observed spectra with template spectra. RVSAO provides the $r$-value defined by \citet{Tonry1979}, which indicates the reliability of the cross-correlation result. We include only galaxies with $r > 4$, consistent with the threshold verified by visual inspection \citep{Geller2014b, Geller2016, Kang2025}. We add a total of 2,032 new redshifts for the A2199 field within $35^\prime$, obtained from \citet{Song2017} and from this study.

We then combine this dataset with that used in \citet{Song2017}, which consists of the redshift data from the SDSS DR17 and that from their MMT/Hectospec survey. The SDSS DR17 provides spectroscopic redshifts for nearly all galaxies down to $m_{r,\mathrm{Petro}} = 17.77$ mag, and some fainter galaxies \citep{Strauss2002}. In total, we obtain 325 redshifts from the SDSS in the A2199 field within $35^\prime$. In addition, we include 384 redshifts from the DESI DR1. We also compile a total of 43 redshifts from the NASA/IPAC Extragalactic Database.

We then perform a visual inspection to check the photometry quality of our A2199 dataset. We select objects with $m_{r,\mathrm{Petro},0}<21$ regardless of colors from the photometric catalog, which are within $35^\prime$ of the center of A2199. We identify extended sources using the SDSS star–galaxy flags (see Section~4.2 of \citealt{Strauss2002}). There are 135 sources wrongly classified as point sources from the images, which are included as extended sources if they have redshift estimates. These constitute only about $\sim5\%$ of the galaxies with spectroscopic redshifts in this field. The impact of galaxy classification ambiguity from the SDSS star–galaxy flags is negligible.

On the other hand, we find some sources that are wrongly identified as extended sources; they turn out to be either saturated stars or artifacts (e.g., spikes, stellar bleeding, fragmented source) identified through visual inspection. Among the 4,063 sources classified as extended sources in the photometric catalog (i.e., those with $m_{r,\mathrm{Petro},0}<21$ and located within $35^\prime$ of the center of A2199), we excluded 83 galaxies based on visual inspection. We also find five sources that have unusually bright Petrosian magnitudes given the appearance in the SDSS images. In such cases, we replace the Petrosian magnitude with the fiber magnitude. We visually compare the brightness of each source with that of nearby objects with valid photometry to ensure accuracy of the magnitude estimates.

The bottom left panel of Figure~\ref{fig:fspec_a2199} shows the spectroscopic completeness across $r$-band magnitude within $30^\prime$. We see a clear improvement in spectroscopic completeness achieved in this study. The vertical black and red dotted lines indicate the magnitude limits where the spectroscopic completeness falls below 50\%, before and after this study, respectively. The completeness limit is extended by 0.6 magnitudes, reaching \( m_{r,\mathrm{Petro},0} = 20.8 \).

The right panel of Figure~\ref{fig:fspec_a2199} shows the two-dimensional spatial distribution of spectroscopic completeness. Completeness remains high in the central region within \( 30^\prime \) but drops rapidly toward the edges. A radius of \( 30^\prime \) corresponds to approximately \( \sim 0.78R_{200} \sim 1.15R_{500} \) of the A2199 cluster (HeCS-omnibus catalog, \citealp{Sohn2020}). This region effectively covers the overdense environment at the core of the cluster. The luminosity function of A2199 is constructed in the central area, where the spectroscopic completeness appears to be high, homogeneous, and isotropic.

\begin{figure}[t]
    \centering
    \includegraphics[width=0.45\textwidth]{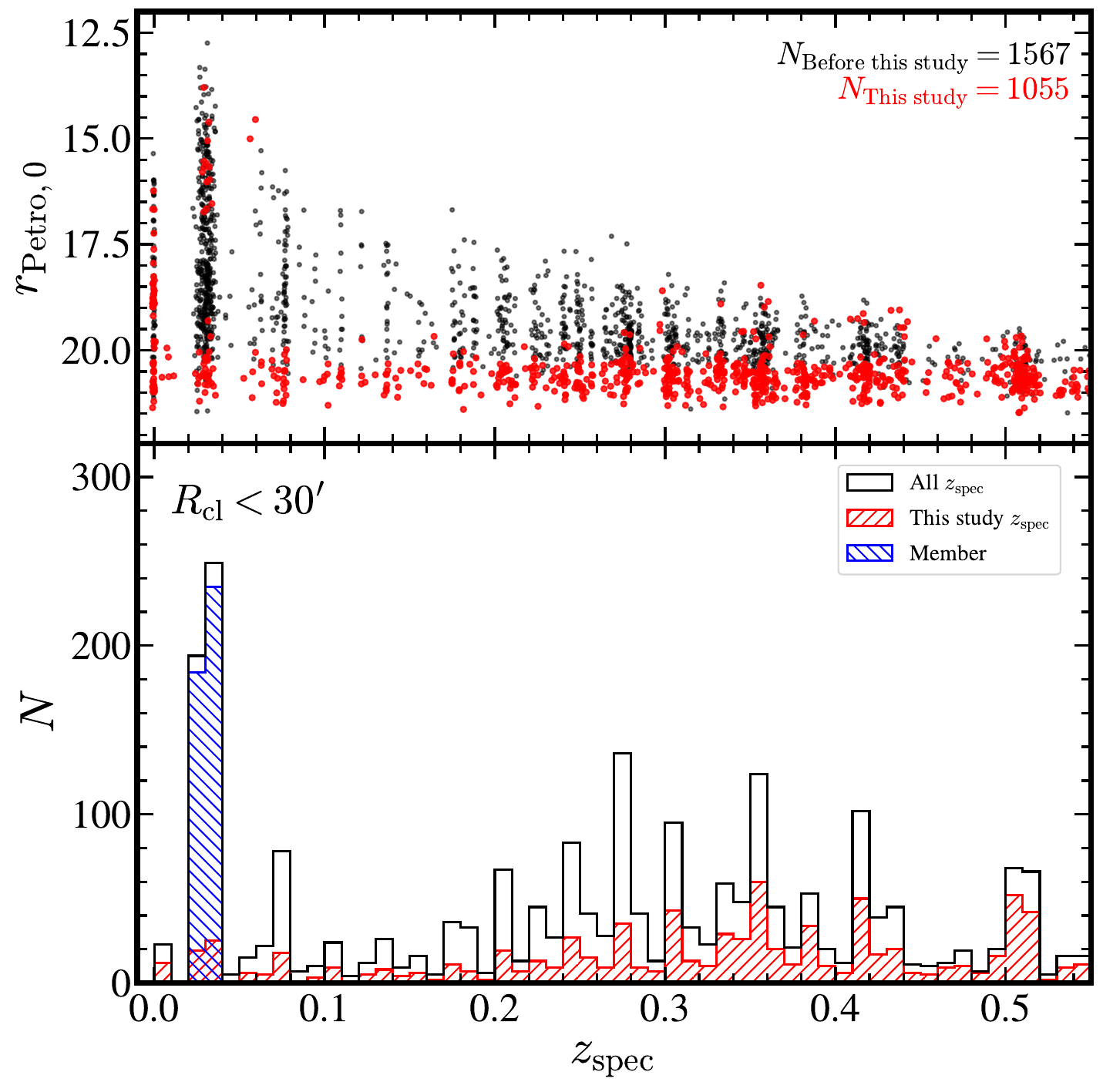}
   \caption{(Top) Distribution of $m_{r,\mathrm{Petro},0}$ as a function of redshift. Red dots indicate the galaxies with redshifts from this study. (Bottom) Redshift histogram of galaxies. The black histogram shows all galaxies in the sample. The blue histogram shows the distribution of member galaxies. Red histogram shows the redshifts from this study.}
    \label{fig:z_number_a2199}
\end{figure}

Figure~\ref{fig:z_number_a2199} shows the redshift distribution of the data compiled in this study. The top panel displays the apparent magnitudes of galaxies as a function of redshift. Black dots represent the number of galaxies with measured redshifts before this study, while red dots indicate galaxies added in this study. We significantly increase the number of galaxies with spectroscopic redshifts in the A2199 field within $35^\prime$ from 1,678 to 2,784 by adding 1,106 new redshifts. Note that the numbers indicated in Figure \ref{fig:z_number_a2199} are for $R_\mathrm{cl}<30^\prime$.

The bottom panel of Figure~\ref{fig:z_number_a2199} shows the number of galaxies across redshift. The blue hatched histogram highlights the member galaxies of A2199 (to be discussed in Section~\ref{sec:cluster_membership} about the cluster member classification), which are concentrated near the cluster redshift (\( z_\mathrm{cl} = 0.03 \)). Our additional spectroscopic data contribute predominantly to the faint galaxy population at relatively higher redshifts. The new spectroscopy from MMT/Hectospec and DESI plays a crucial role in distinguishing interlopers from true cluster members in the A2199 field.

In Table~\ref{tab:galaxy_catalog}, we list all available redshifts in the central region of A2199 within a radius of $35^\prime$. The table provides the SDSS DR17 ObjID, right ascension, declination, $r$-band apparent magnitude, classification flag for extended sources, photometric selection flag, spectroscopic redshift and its error, redshift source, and cluster membership. Note that the classification flags are determined by combining the SDSS star–galaxy flags and the results of our visual inspection.

\begin{table*}[t]
  \centering
  \caption{Redshifts in the field of A2199 within $35^\prime$ from the cluster center.}
  \resizebox{0.95\textwidth}{!}{%
    \begin{tabular}{c c c c c c c c c c c}
      \toprule
      ID & SDSS DR17 ObjID & R.A.($^\circ$) & Decl.($^\circ$) & $r_{0}$(mag) & Flag$_\mathrm{extended}$\textsuperscript{(a)} & Flag$_\mathrm{photo}$\textsuperscript{(b)} & $z$ & $z_\mathrm{err}$\textsuperscript{(c)} & $z$ Source\textsuperscript{(d)} & Member\textsuperscript{(e)} \\
      \midrule
      1 & 1237659325492363706 & 246.411225 & 39.486061 & 21.322 & 0 & 0 & 0.6151 & 0.0002 & 2 & 0 \\
      2 & 1237659325492297918 & 246.414101 & 39.546131 & 15.115 & 1 & 0 & 0.0302 & 0.0001 & 2 & 1 \\
      3 & 1237659325492298799 & 246.416353 & 39.537589 & 20.899 & 1 & 0 & 0.7277 & 0.0001 & 3 & 0 \\
      4 & 1237659325492363724 & 246.421577 & 39.462481 & 20.877 & 0 & 0 & -0.0001 & 0.0001 & 4 & 0 \\
      5 & 1237659325492297794 & 246.424306 & 39.600097 & 13.791 & 1 & 0 & 0.0298 & 0.0001 & 2 & 1 \\
      6 & 1237659325492299063 & 246.430491 & 39.617078 & 19.081 & 1 & 0 & 0.7461 & 0.0001 & 3 & 0 \\
      7 & 1237659325492364609 & 246.436615 & 39.441197 & 21.364 & 1 & 0 & 0.6928 & 0.0001 & 3 & 0 \\
      \bottomrule
    \end{tabular}%
  }
  \captionsetup{justification=raggedright,singlelinecheck=false}
  \caption*{NOTE : A full version of this table is provided in machine-readable format in the online journal. The portion shown here serves as an example of its structure and content.}
  \caption*{(a) (0) Point source, (1) Extended source \\
            (b) (0) Petrosian magnitude, (1) Fiber magnitude \\
             (c) If redshift errors of galaxies from NED are not available or smaller than 0.0001, we set them to 0.0001. Redshifts and uncertainties are reported to 0.0001 precision. \\
             (d) (1) This study, (2) SDSS DR17, (3) DESI DR1, (4) \cite{Albareti2017}, (5) \cite{Song2017}, (6) \cite{Zaritsky2023} \\
             (e) (0) Non-member (1) Member \\
             }
  \label{tab:galaxy_catalog}
\end{table*}

\begin{figure*}[t]
    \centering
    \includegraphics[width=0.8\textwidth]{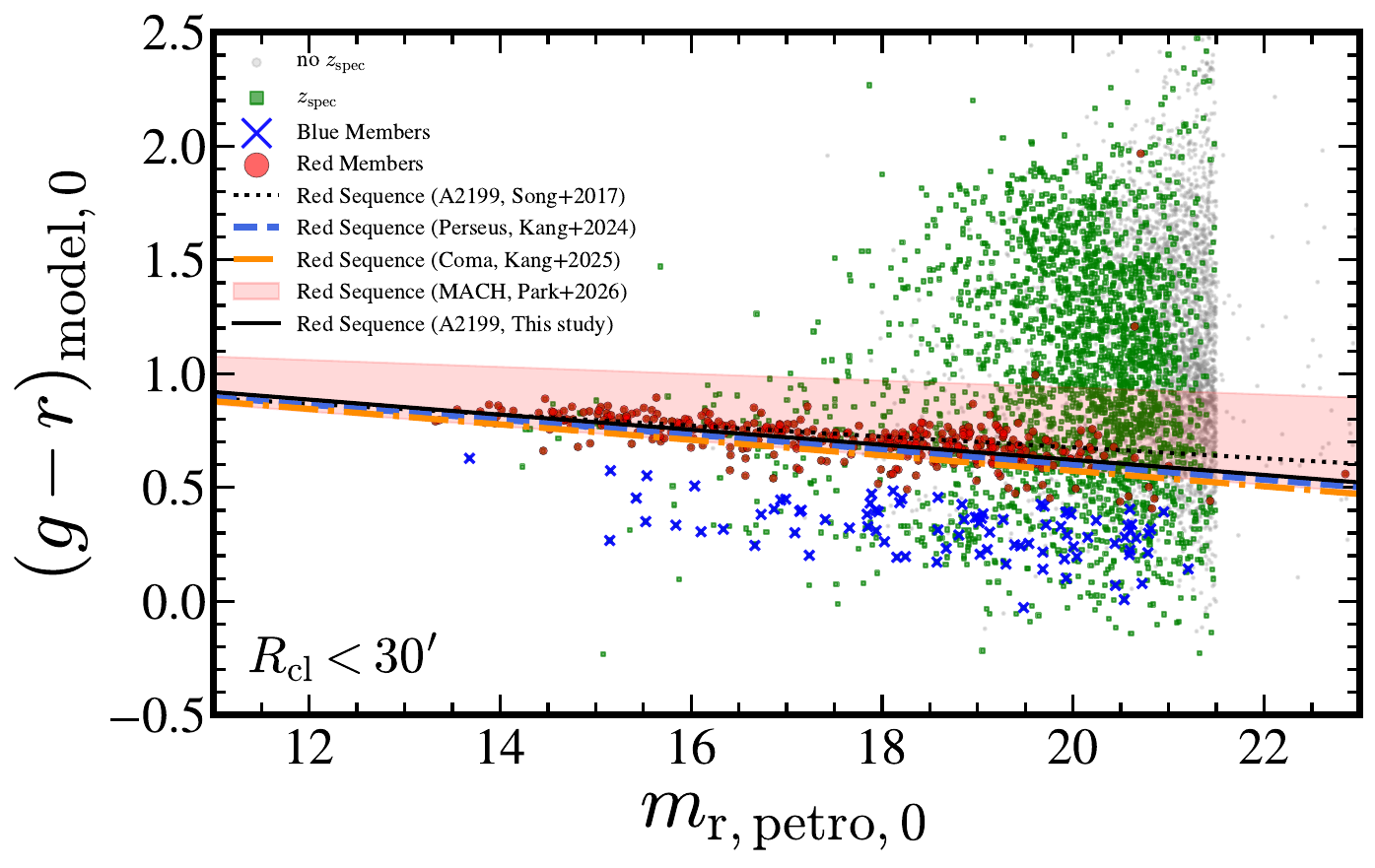}
    \caption{
        $(g-r)_\mathrm{model}$ color vs. $m_{r,\mathrm{Petro},0}$ magnitude diagram of galaxies within $30^\prime$ of the A2199 cluster center. 
        Gray dots show galaxies without spectroscopic redshifts. 
        Green squares indicate galaxies with spectroscopic redshifts. Red circles are red sequence member galaxies and the blue crosses are blue member galaxies. The member galaxies are spectroscopically confirmed. The black solid line shows the red sequence of A2199 members in this study. The black dotted line shows the red sequence of A2199 members from \cite{Song2017}. The blue dashed line shows the red sequence of Perseus members from \cite{Kang2024}. The orange dot-dashed line shows the red sequence of Coma members from \cite{Kang2025}. Red shade shows the range of red sequence of the MACH clusters at $0.07<z<0.11$ \citep{Park2026}}
    \label{fig:cmd_a2199}
\end{figure*}

Figure~\ref{fig:cmd_a2199} shows the color–magnitude diagram (CMD) of galaxies within $30^\prime$ from the cluster center in the A2199 field. Galaxies without spectroscopic redshifts are shown as gray dots, and galaxies with spectroscopic redshifts are shown by colored symbols. We show red member galaxies as red circles, and blue member galaxies as blue crosses.

We apply a linear fit with $3\sigma$ clipping to spectroscopically confirmed member galaxies satisfying $m_{r,\mathrm{Petro},0} < 18$ and $(g - r)_\mathrm{model,0} > 0.5$. We determine the resulting best-fit red sequence as:
\begin{equation}
\begin{split}
 (g - r)_\mathrm{model,0} &= \\ (-0.031 \pm 0.004)\, m_{r,\mathrm{Petro},0}
 & + (1.256 \pm 0.056).
\end{split}
\end{equation}

\noindent 
with a standard deviation of $\sigma = 0.06$. Blue member galaxies are defined as those lying more than $3\sigma$ below the fitted red sequence. Among the 419 member galaxies within $30^\prime$ of the cluster center, 88 (21.0\%) are classified as blue members.

We plot the red sequence derived in this study as a solid black line in Figure~\ref{fig:cmd_a2199}. We observe a well-defined red sequence for A2199 members. For comparison, we display the A2199 red sequence from \citep{Song2017} as a black dotted line. 
We also show the red sequences from two local clusters, Perseus \citep{Kang2024} and Coma \citep{Kang2025} as blue and orange lines, respectively.
We represent the red sequence of MACH clusters \citep{Sohn2020} at $0.07<z<0.11$ \citep{Park2026} as a shaded red region.  The comparative red sequences closely align with the red sequence derived in this study.

\subsection{Cluster Membership} \label{sec:cluster_membership}

Accurate cluster membership is essential for constructing luminosity functions. We use the caustic technique to determine cluster membership. The caustics is the envelope of cluster members in redshift space (radial velocity vs. projected cluster distance), which indicates the escape velocity at a given radius \citep{Diaferio1997, Diaferio1999}. We interpret the caustics as an escape velocity profile at a given radius and consider galaxies lying inside the caustic boundary as members \citep{Diaferio1997, Diaferio1999, Serra2011, Serra2013}. 

$N$-body simulations show that the caustic technique can recover approximately 97\% of true cluster members within $R_{200}$ in systems with more than 200 galaxies \citep{Serra2013}. Numerous spectroscopic surveys have adopted this technique for member selection (e.g., \citealp{Rines2006, Rines2013, Rines2016, Hwang2014, Sohn2017, Sohn2020, Pizzardo2021}).  We identify cluster members using the \texttt{CausticSNUpy} package\footnote{https://github.com/woodykang/CausticSNUpy} \citep{Kang2024}, a Python-based implementation of the method described by \citet{Diaferio1999} and \citet{Serra2011}.

\begin{figure*}[t]
    \centering
    \includegraphics[width=0.7\textwidth]{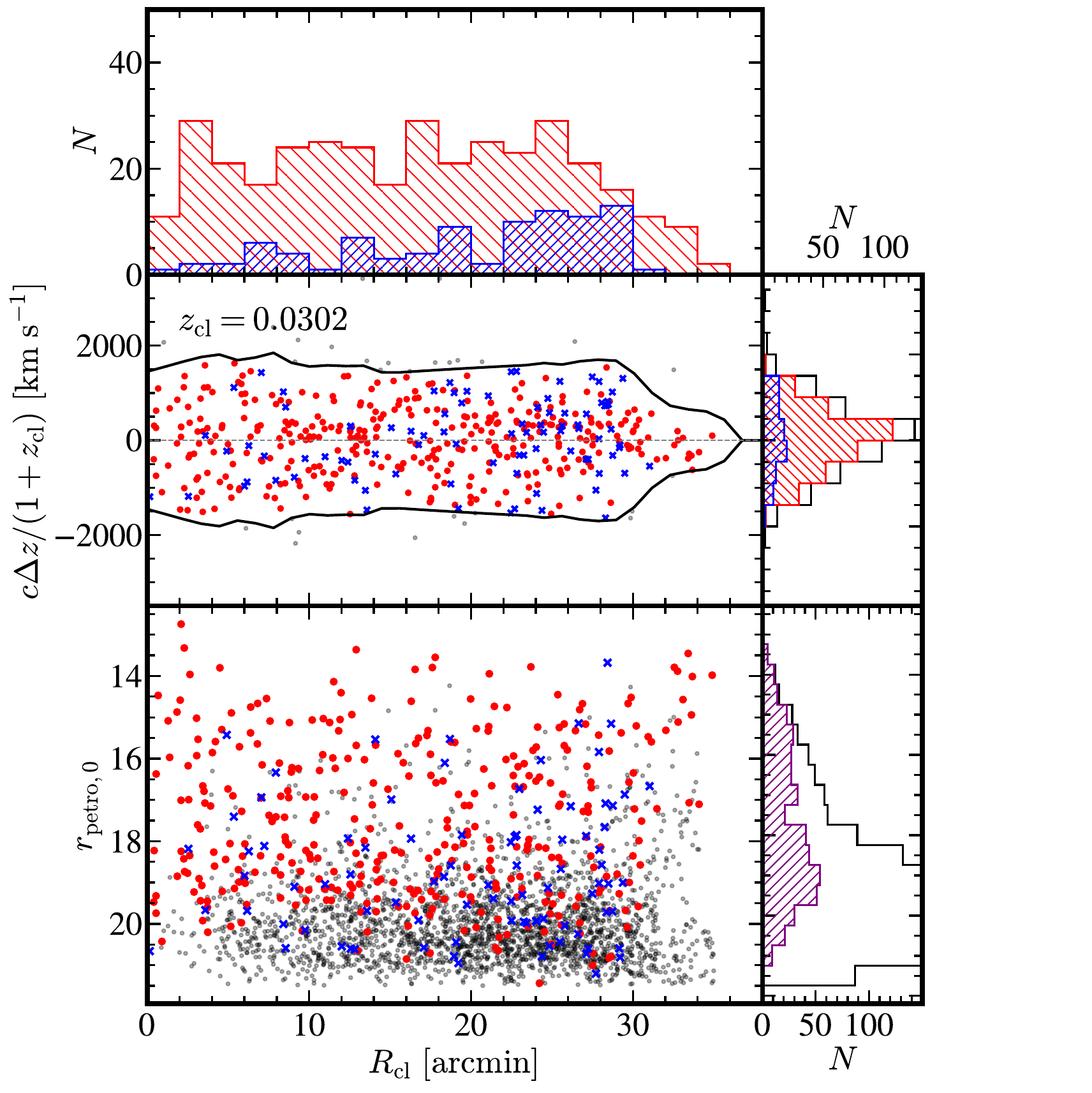}
\caption{(Middle left) Rest-frame cluster-centric velocities of galaxies as a function of projected cluster-centric radius. The caustic profile of A2199 is shown as a black solid line. Red circles and blue crosses indicate red and blue member galaxies, respectively. Black dots represent non-member galaxies. (Middle right) Histograms of rest-frame cluster-centric velocities for galaxies with redshifts in the A2199 field. Red and blue hatched histograms correspond to red and blue member galaxies, respectively. The black solid-line histogram includes all the number of galaxies with measured redshifts in the field. (Top) Histograms of the radial distribution of red and blue member galaxies as a function of cluster-centric distance.
(Bottom left) Apparent $r$-band Petrosian magnitude ($m_{r,\mathrm{Petro},0}$) versus projected cluster-centric distance. Black dots indicate non-member galaxies with measured redshifts (Bottom right) Histogram of $m_{r,\mathrm{Petro},0}$. The purple histogram shows all spectroscopically confirmed member galaxies in A2199, while the black histogram shows all non-member galaxies with measured redshifts within the A2199 field.}

    \label{fig:caustic_a2199}
\end{figure*}

The middle-left panel of Figure ~\ref{fig:caustic_a2199} presents the caustic profile of A2199 in rest-frame velocity versus projected cluster-centric distance space within $35^\prime$. Red and blue crosses represent red and blue member galaxies, respectively. We identify 443 galaxies as cluster members from a sample of 2,784 galaxies with spectroscopic redshifts within $35^\prime$. We note that the spectroscopic completeness of the sample decreases at $r > 30^\prime$, which may potentially influence the caustic profile. However, we find good agreement between the caustic derived from the spectroscopically complete inner sample ($R_\mathrm{cl} < 30^\prime$) and that derived using the extended, spectroscopically incomplete sample ($R_\mathrm{cl}<120^\prime$; comparison not shown here). This agreement demonstrates that the caustic determination within the central region is robust against spectroscopic incompleteness at larger radii.

The top-left and middle-right panels display histograms of galaxy distributions as a function of projected clustercentric distances and radial velocities of galaxies, respectively. The red galaxy population clearly dominates the central region. As we move inward to $R_\mathrm{cl} < 10^\prime$, the fraction of blue member galaxies further declines. This is primarily due to a decrease in the number of blue galaxies, while the number of red galaxies remains approximately constant.

The bottom-left panel shows the distribution of apparent Petrosian magnitudes ($m_{r,\mathrm{Petro},0}$) as a function of projected cluster-centric distance. The bottom-right panel displays the histogram of $m_{r,\mathrm{Petro},0}$. Most objects are distributed within $R_\mathrm{cl} < 30^\prime$ and  $m_{r,\mathrm{Petro},0}<21$, where spectroscopic completeness is high, as seen in Figure~\ref{fig:fspec_a2199}. Member galaxies of A2199 are typically brighter than non-member (interloper) galaxies. 

\section{Results : Luminosity Function of A2199} \label{sec:luminosity_function_a2199}

The luminosity function of A2199 is constructed by counting member galaxies as a function of magnitude. However, these counts must be corrected for both photometric and spectroscopic incompleteness. Photometric incompleteness is primarily caused by low surface brightness objects. To account for the incompleteness in the SDSS data used in this study, we adopt the correction factor $C_\mathrm{SB}$ provided by \citet[see their Figure~6]{Blanton2005}. Spectroscopic incompleteness can be corrected using the spectroscopic completeness fraction ($f_\mathrm{spec}$) shown in Figure~\ref{fig:fspec_a2199}, defined as the number of galaxies with redshifts among all galaxies in a given magnitude bin. The corrected luminosity function is then computed by weighting each galaxy by $(1/f_\mathrm{spec})\cdot C_\mathrm{SB}$. 

We note that our spectroscopic targets are selected in a flux-limited manner, without explicit color preselection. As a result, the spectroscopic completeness does not exhibit a strong dependence on galaxy color, and applying separate completeness corrections for red and blue galaxies yields luminosity function parameters consistent within the uncertainties. In addition, although our analysis is restricted to $R_{\rm cl}<0.78R_{200}$, tests using simulated cluster samples indicate that the exclusion of faint blue galaxies in the outskirts does not significantly affect the faint-end slope measured within this radius.

\begin{figure*}[t]
    \centering
    \includegraphics[width=0.85\textwidth]{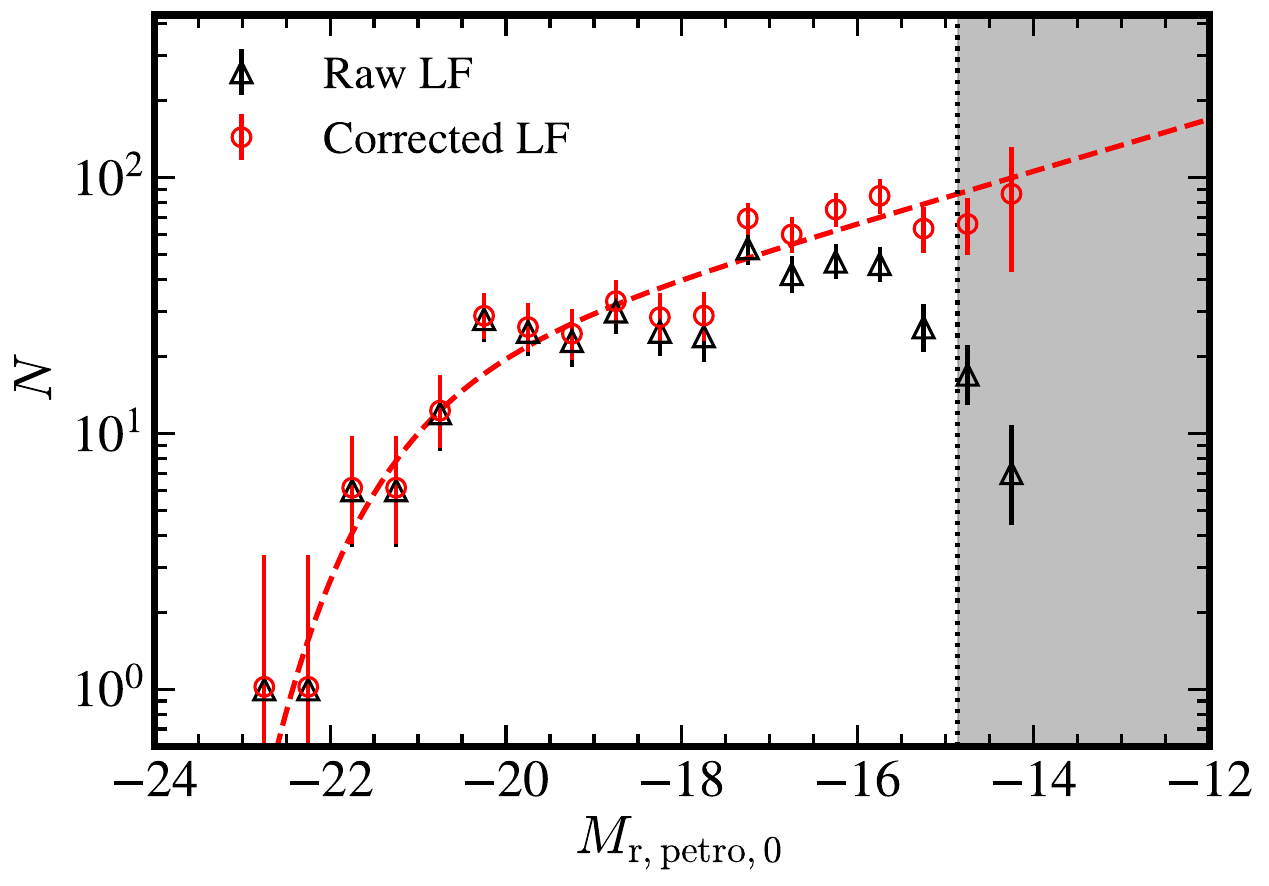}
    \caption{Luminosity function of A2199. The black triangles represent the raw luminosity function. The red circles indicate the luminosity function corrected for data incompleteness. Error bars represent Poisson errors. The red dashed line shows the Schechter function fit. The gray shaded region indicates the magnitude range where the spectroscopic completeness is below $f_\mathrm{spec} < 0.5$.}
    \label{fig:luminosity_function_a2199}
\end{figure*}

Figure~\ref{fig:luminosity_function_a2199} shows the luminosity function of A2199. The black triangle shows the simple number counts of member galaxies in our A2199 catalog (i.e., the ``raw'' luminosity function). The red circles indicate the luminosity function after correcting for data incompleteness. The black dotted vertical line marks the absolute magnitude corresponding to the spectroscopic completeness limit—defined as the apparent magnitude at which the spectroscopic completeness ($f_\mathrm{spec}$) falls below 50\%—as shown in left panel of Figure~\ref{fig:fspec_a2199}.

The luminosity function is commonly modeled using the Schechter function \citep{Schechter1976}. The Schechter function takes the form of a power law for low luminosities and transitions to an exponential decline for high luminosities as follows:

\begin{equation}
    \phi(M) = \phi_0 \, 10^{0.4(1+\alpha)(M^* - M)} \exp\left[-10^{0.4(M^* - M)}\right]
\end{equation}

\noindent Three parameters define the function: where $M$ is a magnitude, $\phi^*$ is a normalization constant, $M^*$ is the characteristic magnitude marking the transition point (often referred to as the “knee magnitude” of the luminosity function), and $\alpha$ is the faint-end slope that describes how rapidly the number of galaxies changes at lower luminosities. We quantify the faint-end slope, $\alpha$, of the cluster luminosity function by fitting Schechter function. 

We fit the Schechter function over the magnitude range where the spectroscopic completeness exceeds $f_\mathrm{spec} > 0.5$, excluding the fainter region below this limit, which we consider incomplete. We perform the fitting using a Markov Chain Monte Carlo (MCMC) method implemented with the \texttt{emcee} package \citep{ForemanMackey2013}. The best-fit result, with $M^*=-21.30 \pm {0.27}$ and  $\alpha=-1.23 \pm {0.05}$, is indicated by the red dashed line in Figure~\ref{fig:luminosity_function_a2199}.

\begin{figure*}[t]
    \centering
    \includegraphics[width=0.8\textwidth]{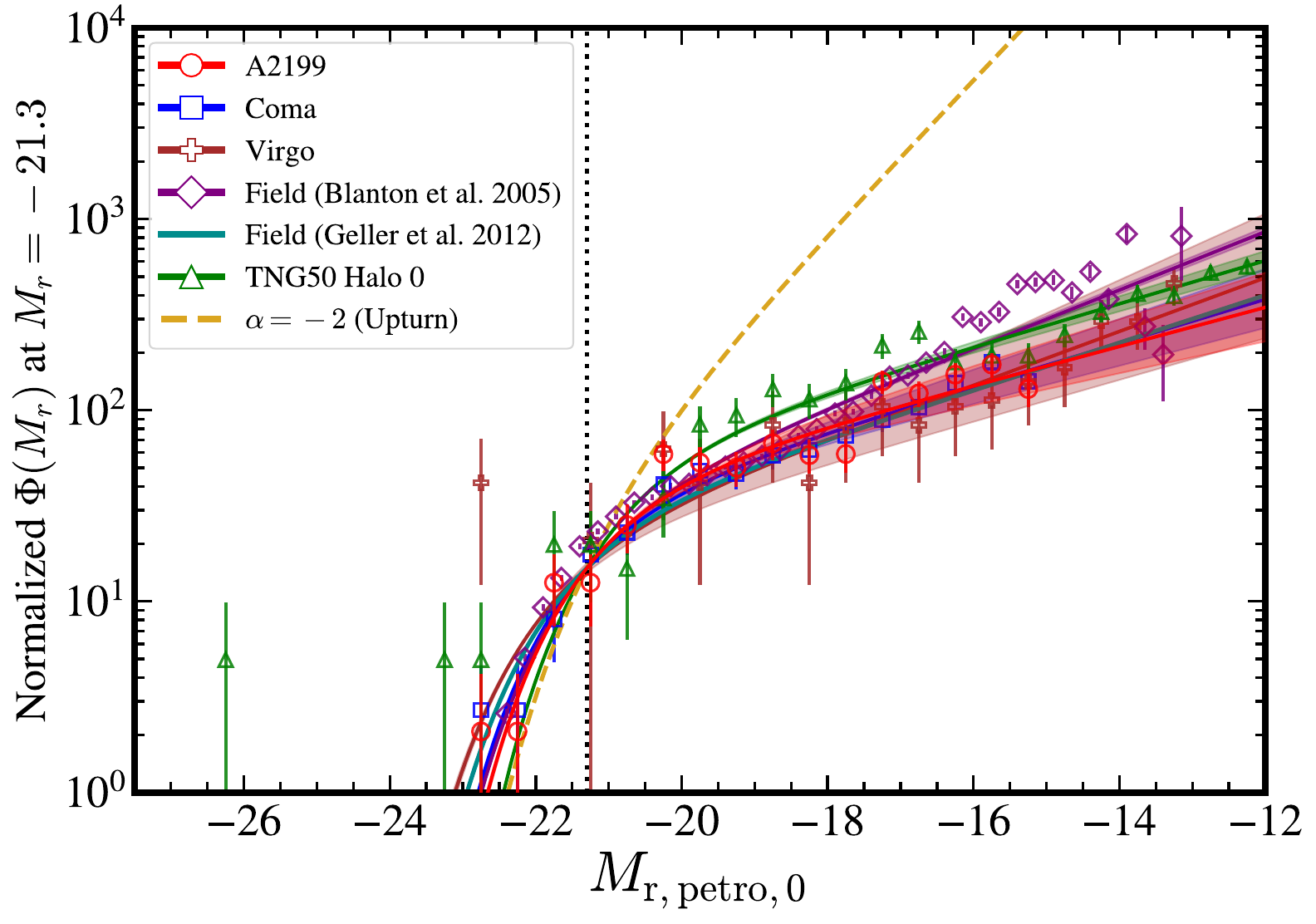}
\caption{Comparison of the luminosity function of A2199 with those of the Coma and Virgo clusters, the local field, and the TNG50 simulations. The black dotted line marks the characteristic magnitude from the A2199 luminosity function. All luminosity functions are shifted to intersect at the characteristic magnitude of A2199 luminosity function. Solid lines represent the Schechter function fits, with shaded regions indicating the slope ranges derived from uncertainties in the fitting. The gold dashed line represents the luminosity function with a steep faint-end slope ($\alpha = -2$).}
    \label{fig:luminosity_function_total_comparison}
\end{figure*}

\section{Discussion} \label{sec:discussion}

The observed faint-end slopes of field luminosity functions are typically in the range $-1.6 < \alpha <-1.1$ \citep{Efstathiou1988, Liu2008}.  Theoretically, processes such as tidal interactions and shielding from ultraviolet radiation in galaxy clusters can respectively induce and preserve dwarf galaxies, leading to an excess population of low-luminosity systems. This excess may steepen the faint-end slope of the luminosity function \citep{Barnes1992, Bekki2001, Tully2002, Benson2003, Popesso2006}.

\citet{Rines2008} derived the galaxy luminosity function of A2199 down to $r = 20$ using MMT/Hectospec redshift data for galaxies within $R_\mathrm{cl} < 30^\prime$. They reported Schechter parameters of $M^* = -21.11^{+0.21}_{-0.25}$ and a faint-end slope of $\alpha = -1.13^{+0.07}_{-0.06}$, consistent with typical values found in other galaxy clusters. However, their spectroscopic completeness in the central region of A2199 was limited to 38\% for the galaxies at $m_{r,\mathrm{Petro},0}<20.5$, which may have limited the robustness of their luminosity function at the faint end.

\citet{Song2017} significantly improved the spectroscopic completeness to 77\% in the same magnitude range. They compiled a total of 1,624 redshifts, including 775 newly added measurements. They extended the A2199 luminosity function to $r < 20.5$ using reliable redshift identifications. Their Schechter fit yielded parameters of $M^* = -21.30 \pm 0.37$ and $\alpha = -1.10 \pm 0.06$. Although their deeper data allowed them to probe fainter galaxies, they still found no evidence for a faint-end upturn in the luminosity function at $M_r < -15$.

In this study, we compile additional redshift measurements from MMT and DESI, extending the A2199 luminosity function to \( r \simeq 20.8 \). The luminosity function reaches down to \( M_r = -14 .5\). By nearly doubling the number of galaxies with measured redshifts, we push 0.6\,mag deeper than previous work, yet we still detect no evidence for a faint‑end upturn in luminosity function ($\alpha=-1.23\pm0.05$). We now compare the A2199 luminosity function to those derived for other samples.

\subsection{Comparison with other samples} \label{sec:compare_samples}

\subsubsection{Observed Local Clusters: Coma and Virgo}

Coma cluster is one of the most massive galaxy clusters in the local universe. It is located at redshift $z \sim 0.023$, with a mass of $M_{200} \sim 10^{14.8}~\mathrm{M_\odot}$ and a characteristic radius of $R_{200} \sim 1.76~\mathrm{Mpc}$ (HeCS-omnibus catalog, \citealp{Sohn2020}). Coma shares similar physical properties with A2199, which is located at $z \sim 0.031$, has a mass of $M_{200} \sim 10^{14.5}~\mathrm{M_\odot}$, and a radius of $R_{200} \sim 1.42~\mathrm{Mpc}$ (HeCS-omnibus catalog, \citealp{Sohn2020}).

We use the spectroscopic sample of Coma galaxies of \citet{Kang2025} to construct luminosity function in the central region within $30^\prime$. We follow the same methodology described in Section~\ref{sec:luminosity_function_a2199} to construct luminosity function. The Coma luminosity function is complete down to $M_r = -15$. From the Schechter function fit, we obtain a faint-end slope of $\alpha = -1.29 \pm {0.04}$ and a characteristic magnitude of $M^* = -21.57^{+0.24}_{-0.23}$.

The Virgo cluster has a dominant mass concentration in the local universe \citep{Hong2021}, with a dynamical mass of $M_{200} \sim 5.5 \times 10^{14}\, \rm M_\odot$ \citep{Durrell2014}. While the luminosity function of the Virgo cluster is presented in both \citet{Ferrarese2016} and \citet{Ferrarese2020}, the latter provides a catalog of galaxies in the core of the Virgo cluster ($\sim 3.71\,\deg^2$, $\sim 0.3\,\text{Mpc}^2$), including membership classifications. Based on this catalog, we reconstruct the Virgo cluster luminosity function using galaxies identified as certain and likely members.

\citet{Ferrarese2020} provided photometry in the $ugriz$ bands from the MegaCam instrument on the Canada-France-Hawaii Telescope (CFHT). To ensure consistency with our SDSS-based analysis, we apply the CFHT-to-SDSS filter transformation provided by the MegaPipe documentation\footnote{\url{https://www.cadc-ccda.hia-iha.nrc-cnrc.gc.ca/en/megapipe/docs/filtold.html}}. For Galactic foreground extinction correction, we use SDSS values when available, and for galaxies lacking SDSS photometry, we apply the mean extinction value derived from member galaxies with SDSS data.

For K-corrections, we follow the same procedure described in Section~\ref{sec:photdat}. Because there are only 85 galaxies in Virgo with measured redshifts, we apply a cluster-centric velocity offset of $c\Delta z \sim 1100~\mathrm{km/s}$ to perform K-corrections for the remaining galaxies. 

The reconstructed luminosity function for Virgo is complete down to $M_r = -12$. Although the completeness definition differs from our approach, \citet{Ferrarese2016} reported a 50\% photometric completeness limit at $M_r \sim -10$. We use all member galaxies within $R_\mathrm{cl}<90^\prime$ ($\sim 0.432 \,\mathrm{Mpc}$) to fit the Schechter function.

From our SDSS $r$-band luminosity function of the Virgo cluster, we obtain Schechter parameters of $M^* = -22.18^{+2.16}_{-1.67}$ and $\alpha = -1.36^{+0.08}_{-0.09}$, consistent with the results reported in \citet{Ferrarese2016}. The characteristic magnitude appears brighter and more uncertain than in A2199 and Coma, which is probably because of presence of many bright galaxies in the Virgo core.

To better compare the faint-end slope of Virgo with the other clusters, we fix the characteristic magnitude to $M^* = -21.35$ that is the average value derived from the A2199 and Coma luminosity functions. With this constraint, the faint-end slope remains $\alpha=-1.35\pm0.08$.

\subsubsection{Observed Local Field Galaxies}

To compare LFs of different environments, we bring the $r$-band luminosity function of field galaxies. \citet{Blanton2005} construct field luminosity function using the SDSS DR2 galaxies at $z<0.05$ ($10<d_\mathrm{com}/h^{-1}\text{Mpc}<150$). The field luminosity function is constructed for the magnitude range down to $M_r - 5 \log h \sim -12.5$ (approximately $M_r \sim -13.2$). \citet{Blanton2005} reported a faint-end slope of $\alpha \sim -1.52$, indicating a steeper luminosity function in the field than in typical clusters. Tables 2 and 5 in \cite{Blanton2005} provide the $r$-band luminosity function data points, associated uncertainties, and the best-fit double Schechter function parameters. They report a faint-end slope of $\alpha=-1.38 \pm 0.01$ and a characteristic magnitude of $M_r^*=-21.56 \pm 0.01$. We adopt their field luminosity function for comparison with our results from A2199.

\citet{Geller2012} derived the $R$-band field galaxy luminosity function from the SHELS redshift survey within the redshift range \(0.02 \le z \le 0.1\). SHELS is a spectroscopically deep, complete survey that includes low–surface‑brightness galaxies \citep{Geller2014b, Geller2016}. It is 96\% spectroscopically complete to \(R = 20.6\). SHELS reaches an absolute‑magnitude limit of \(M_R = -13.3 + 5\log h \simeq -14.08\). They obtain a faint‑end slope of \(\alpha \simeq -1.31\) and a characteristic magnitude of \(M_R^* - 5\log h \simeq -21.32\). We correct the characteristic magnitude to the $r$-band using Equation 1 of \citet{Geller2016}, assuming a typical color of $g - r \approx 1$ for the SHELS galaxies (see Figure 7 of \citet{Geller2012}), and thereby obtain a corresponding $r$-band characteristic magnitude of $M_r^* \approx -21.80$. These values are consistent with the faint-end slope found in cluster luminosity functions and likewise show no evidence for a faint‑end upturn in luminosity functions.

\subsubsection{Simulated Local Cluster: TNG50-1}

Cluster luminosity functions derived from observations are limited by the survey depth and uncertainties in membership determination. These limitations in observational luminosity functions hinder the complete characterization at the faint-end in cluster luminosity function. In contrast, high-resolution cosmological hydrodynamic simulations provide robust alternatives for luminosity function studies by offering complete membership information and deeper luminosity limits. 

The IllustrisTNG project is a suite of state-of-the-art cosmological magnetohydrodynamical simulations \citep{Nelson2019, Pillepich2019}. TNG50 is the highest-resolution run in this suite, simulating a 50~Mpc box with a baryonic mass resolution of $\sim 8.5 \times 10^4~\mathrm{M_\odot}$ and spatial resolution down to $\sim 100~\mathrm{pc}$. The high resolution of TNG50 enables accurate modeling of galaxy structure and luminosity, allowing reliable predictions of the luminosity function in cluster environments. 

The baryonic particle resolution of TNG50 is $M_* \sim 8.5 \times 10^4 \, \mathrm{M_\odot}$. To ensure that the luminosity function is not affected by the resolution limit of TNG50, we adopt a lower limit of our luminosity function of TNG50 corresponding to halos with more than 100 particles. This corresponds to $M_* = 8.5 \times 10^6 \, \mathrm{M_\odot}$, which is equivalent to an SDSS $r$-band absolute magnitude of $M_r \sim -12.0$. We extend the cluster luminosity function to depths beyond those accessible with current observations.

We use Snapshot 96 of TNG50, corresponding to $z \sim 0.03$, to match the redshifts of A2199 and Coma. This snapshot contains only one massive halo with $M_{200} \geq 10^{14}~\mathrm{M_\odot}$, which we select as our comparison target. The selected halo has $M_{200} \sim 10^{14.25}~\mathrm{M_\odot}$ and $R_{200} \sim 1.21~\mathrm{Mpc}$, closely matching the physical properties of A2199. As there is only one simulated cluster to be compared, the comparison between observations and simulations here should be considered in a qualitative sense.

For each galaxy in this halo, SDSS $griz$-band absolute magnitudes are provided as the  \texttt{SubhaloStellarPhotometrics} field, which sums the luminosities of all stellar particles. Using the $r$-band magnitudes, we construct the luminosity function of the TNG50 cluster. Because simulations provide complete information, we do not apply completeness corrections. 

TNG50 simulation provides total magnitudes to construct the luminosity function, whereas the luminosity functions from observation are mostly constructed from SDSS $r$-band Petrosian magnitudes. Total magnitudes are systematically brighter; therefore, TNG50 galaxies could appear slightly more luminous than their observational counterparts. In practice, the mean offset between the SDSS model magnitude and the Petrosian magnitude for A2199 members is only $\langle m_{r,\mathrm{model}} - m_{r,\mathrm{Petro}}\rangle \simeq -0.05$, indicating that the difference is negligible. Applying a Petrosian aperture to the simulated galaxies would therefore shift the luminosity function by less than 0.05\,mag across all luminosity bins. This small shift of the TNG50 luminosity function would have little effect on the overall shape or the faint-end slope.

We fit a Schechter function to the TNG50 luminosity function. For the full faint range down to $M_r \leq -12$, we obtain $\alpha = -1.28 \pm {0.01}$ and $M^* = -21.01^{+0.31}_{-0.40}$. When restricting the fit to $M_r \leq -14$, to match the observational limit of A2199 and Coma, we find $\alpha = -1.14 \pm {0.05}$ and $M^* = -20.19^{+0.30}_{-0.35}$. This result shows that, under observational constraints, the simulated luminosity function also appears shallower at the faint end.

\subsection{No upturn of the faint-end slope of cluster galaxy luminosity function} \label{sec:compare_slope}

To investigate the presence of upturn in faint-end slope of luminosity function of clusters, we compare the luminosity functions derived from observed and simulated clusters, as well as from the field. Figure~\ref{fig:luminosity_function_total_comparison} shows the luminosity functions derived in this study, including those for A2199, Coma, Virgo, TNG50, and the field. For the field luminosity function from \citet{Geller2012}, we use the best-fit Schechter parameters derived by them since the individual data points of the field luminosity function were not provided.

To focus solely on differences in the faint-end slope, we normalize all luminosity functions. Because it is difficult to define the occupied volume of observed clusters owing to the ambiguity of cluster boundaries and projection effects in observations, it is hard to normalize each cluster luminosity function by volume. Instead, we shift the luminosity functions so that they intersect at the characteristic magnitude of the A2199 luminosity function. This normalization allows us to compare the relative shapes of the luminosity functions, independent of differences in the total number of galaxies in each sample.

The black dotted vertical line indicates the characteristic magnitude of the A2199 luminosity function, which serves as a reference magnitude for normalization. Solid lines represent the best-fit Schechter functions, while shaded regions show the uncertainty in the faint-end slope for each fit. A yellow dashed line illustrates a reference Schechter function, sharing the same $M^*$ value as the A2199 luminosity function, but with a steep faint-end slope of $\alpha = -2.0$, for visual comparison. We summarize the normalized luminosity function data in Figure~\ref{fig:luminosity_function_total_comparison} in Table~\ref{tab:normalized_lf_data}. 

\begin{table}[t] 
  \centering
  \caption{Normalized luminosity functions in different environments.}
  \resizebox{0.45\textwidth}{!}{%
    \begin{tabular}{c c c c}
      \toprule
      Sample ID\textsuperscript{(a)} & $M_r$\textsuperscript{(b)} & $\Phi$\textsuperscript{(c)} & $\sigma_{\Phi}$\textsuperscript{(d)} \\
      & (mag) & ($h^3 \mathrm{Mpc}^{-3} \mathrm{mag}^{-1}$) & ($h^3 \mathrm{Mpc}^{-3} \mathrm{mag}^{-1}$) \\
      \midrule
      1 & -22.75 & 2.1 & 2.1 \\
      1 & -22.25 & 2.1 & 2.1 \\
      1 & -21.75 & 12.5 & 5.1 \\
      1 & -21.25 & 12.5 & 5.1 \\
      1 & -20.75 & 25.1 & 7.2 \\
      \bottomrule
    \end{tabular}%
  }
  \captionsetup{justification=raggedright,singlelinecheck=false}
  \caption*{NOTE : A full version of this table is provided in machine-readable format in the online journal. The portion shown here serves as an example of its structure and content.}
  \caption*{
    (a) 1 = A2199 (This study), 2 = Coma \citep{Kang2025}, 3 = Virgo \citep{Ferrarese2020}, 4 = Field \citep{Blanton2005}, 5 = TNG50 (This study) \\
    (b) Absolute magnitudes computed assuming $\Lambda$CDM cosmology with $H_0=70$ km/s/Mpc, $\Omega_m=0.3$, $\Omega_\Lambda=0.7$. \\
    (c) $\Phi$ values are normalized to the luminosity function of A2199 at $M_r = -21.3$. \\
    (d) Errors in $\Phi$ are from Poisson statistics. \\
  }
  \label{tab:normalized_lf_data}
\end{table}

\begin{table}[t]
  \captionsetup{justification=raggedright, singlelinecheck=false}
  \centering
    \caption{Schechter fit parameters for luminosity functions in different environments.}
  \resizebox{0.45\textwidth}{!}{%
    \begin{tabular}{lcc}
      \toprule
      Dataset& $\alpha$ & $M^*_r$ \\
      \midrule
      A2199         & $-1.23^{+0.05}_{-0.05}$ & $-21.30^{+0.27}_{-0.27}$ \\
      Coma \citep{Kang2025}         & $-1.29^{+0.04}_{-0.04}$ & $-21.57^{+0.24}_{-0.23}$ \\
      Virgo \citep{Ferrarese2020}        & $-1.36^{+0.08}_{-0.09}$ & $-22.18^{+2.16}_{-1.67}$\\
      TNG50  & $-1.28^{+0.01}_{-0.01}$ & $-21.01^{+0.31}_{-0.40}$ \\
      Field \citep{Blanton2005} & $-1.38^{+0.01}_{-0.01}$ & $-21.56^{+0.01}_{-0.01}$ \\
      Field \citep{Geller2012}\textsuperscript{$\dagger$} & $-1.31^{+0.04}_{-0.04}$ & $-21.80^{+0.30}_{-0.30}$\\

      \bottomrule
    \end{tabular}%
  }
  \caption*{$\dagger$ It should be noted that the LF is originally defined in the $R$ band rather than the $r$ band. We do not adjust the faint-end slope, as the magnitude difference between $R-$ and $r-$ bands is nearly constant (see equation 1 of \citet{Geller2016}), but we correct the magnitude using the same equation, assuming a typical $g-r\approx 1$ for the SHELS galaxies (see Figure 7 of \citet{Geller2012}).}
  \label{tab:luminosity_function_params}
\end{table}

\begin{figure}[t]
    \centering
    \includegraphics[width=0.5\textwidth]{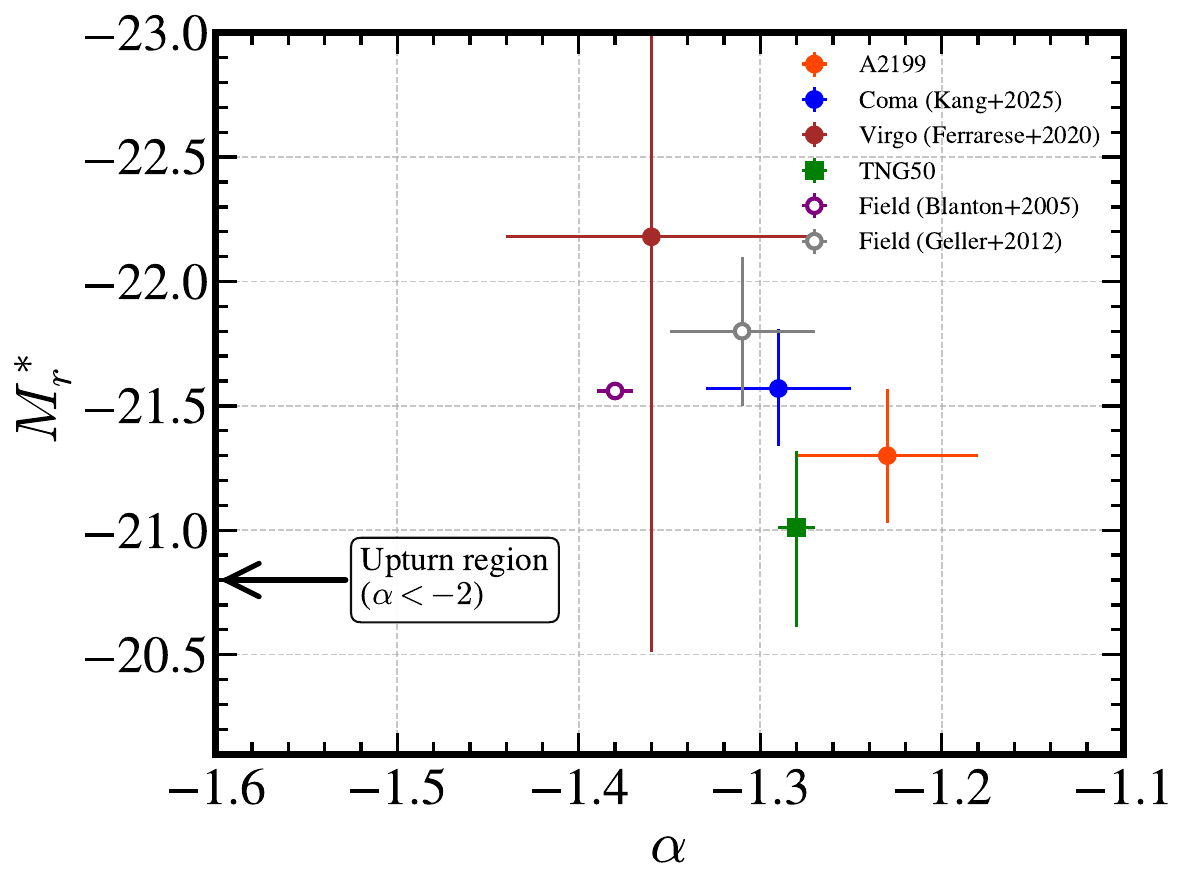}
    \caption{Faint-end slope ($\alpha$) versus characteristic magnitude ($M^*$) of luminosity functions. Error bars indicate parameter uncertainties derived from the MCMC fitting. Open circles are for the field luminosity functions. Filled circles represent cluster luminosity function parameters, and squares show simulation luminosity function parameters.}
    \label{fig:luminosity_function_total_param_comparison}
\end{figure}

Table~\ref{tab:luminosity_function_params} provides a summary of the best-fit Schechter parameters for all luminosity functions analyzed in this study. We find that the faint-end slopes are broadly consistent across the cluster environments of A2199, Coma, and Virgo, as well as in the TNG50 simulation with no indication of a faint-end upturn—i.e., no evidence for a steep slope such as $\alpha \sim -2.0$. This result is further illustrated in Figure~\ref{fig:luminosity_function_total_param_comparison}, which presents the best-fit Schechter parameters along with their associated uncertainties for all luminosity functions. Note that the faint-end slope of the local field luminosity function derived from the \citet{Blanton2005} data differs from the value reported in the literature because we fit a single Schechter function for consistency with the other luminosity functions, whereas the published value is based on a double Schechter fit. 

It should be noted that the member selection for galaxy luminosity function of Virgo \citep{Ferrarese2020} differs from that of A2199 and Coma. The member galaxies in Virgo were selected based on the combination of spectroscopic redshifts and photometric properties including surface brightness, morphology, and size. These criteria account for the low surface brightness and extended appearance typical of Virgo members at a given magnitude, especially among dwarf galaxies that show smooth, irregular structures and lack prominent bulge-disk features. TNG50 assigns membership using a Friends-of-Friends (FoF) algorithm, which may include unbound galaxies and thus lead to systematic differences in the luminosity function.

The luminosity function for the cluster in TNG50 simulation shows an unusually bright galaxy population with $M_r < -24$, significantly brighter than the brightest cluster galaxies (BCGs) observed in real clusters. This is due to the inclusion of "fuzz"—stellar particles bound to the group but not to any specific subhalo—in the total light of the central object. As discussed by \citet{Ahvazi2024}, the total stellar mass of the central subhalo in simulations can be overestimated when no aperture limit is applied, making it appear more massive than observed BCGs. Nevertheless, this effect primarily impacts the bright end of the luminosity function and does not significantly affect the faint-end slope.

We find no evidence for an upturn in either cluster or field environments within the local Universe down to $M_r \simeq -14.5$, and its presence is not supported by any spectroscopic sample currently available. Reports of such an upturn in previous studies are most likely artifacts of background-galaxy contamination in analyses relying solely on photometric data. The absence of a faint-end upturn in our luminosity function suggests that environmental processes in clusters do not trigger the formation or survival of low-mass galaxies.

\section{Conclusion} \label{sec:conclusion}

In this study, we have constructed and analyzed the galaxy luminosity function of the nearby galaxy cluster A2199 down to $M_r \sim -14.5$ using deep spectroscopic data. We provide the deep spectroscopic catalog of A2199 within $35^\prime$. We significantly increase the number of spectroscopic redshifts in the A2199 field within $35^\prime$, nearly doubling it from 1,678 to 2,784.

The spectroscopic completeness magnitude limit, where the spectroscopic completeness is 50\%, reaches $m_{r,\mathrm{Petro},0} = 20.8$. We achieve deep and spatially uniform spectroscopic completeness across the central A2199 field within $35^\prime$. We identify 443 members of the A2199 cluster using the caustic technique based on dense spectroscopy.

Our high-quality dataset provides a reliable measurement of the A2199 luminosity function. Our primary goal was to robustly measure the faint-end slope of the luminosity function, which previous studies have suggested may show a steep upturn indicative of enhanced dwarf galaxy populations in dense cluster environments.

We summarize our results as follows:
\begin{itemize}
  \item The luminosity function of A2199 yields a characteristic magnitude of $M^{*} = -21.30 \pm {0.27}$ and a faint-end slope of $\alpha = -1.23 \pm {0.05}$. We find no evidence for a steep upturn at the faint end, such as a slope approaching $\alpha \sim -2$.
  
  \item A comparison with luminosity functions from other nearby clusters shows that the faint-end slopes of Coma ($\alpha = -1.29 \pm {0.04}$) and Virgo ($\alpha = -1.36^{+0.08}_{-0.09}$) are consistent with that of A2199, reinforcing the absence of a pronounced faint-end upturn in local clusters.
    
  \item The local field luminosity function, based on the \citet{Blanton2005} data, also shows a shallow faint-end slope ($\alpha = -1.381^{+0.007}_{-0.006}$). The field luminosity function from \citet{Geller2012} reports $\alpha \sim -1.31$. We find no evidence of a faint-end upturn in the field luminosity function, showing similar behavior to that seen in cluster environments.
    
  \item The cluster luminosity function extracted from the TNG50 simulation, which extends to fainter luminosities, also shows a similarly shallow faint-end slope ($\alpha = -1.28 \pm {0.01}$), in good agreement with our observational results.
  
  \item Our analysis reveals no evidence for a steep faint-end slope either in spectroscopic observations or in a simulation. The faint-end upturn sometimes reported in the galaxy luminosity function is therefore most likely an artifact of background-galaxy contamination in studies that rely exclusively on photometric data. Down to $M_r \simeq -14.5$, we detect no such upturn in either cluster or field environments within the local Universe. This absence indicates that environmental processes in clusters do not trigger the formation or survival of low-mass galaxies.

\end{itemize}

%%% ACKNOWLEDGMENTS (IF ANY) %%%%%%%%%%%%%%%%%%%%%%%%%%%%%%%%%%%%%%%%

\acknowledgments
HSH acknowledges the support of the National Research Foundation of Korea (NRF) grant funded by the Korea government (MSIT), NRF-2021R1A2C1094577, and Hyunsong Educational \& Cultural Foundation.
HSong acknowledges the support of the National Research Foundation of Korea (NRF) grant funded by the Korea government (MSIT), NRF-2022M3K3A1093827, and Global - Learning \& Academic research institution for Master’s·PhD students, and Postdocs (G-LAMP) Program of the National Research Foundation of Korea (NRF) grant funded by the Ministry of Education (No. RS-2025-25442707). 

%%% APPENDICES (IF ANY) %%%%%%%%%%%%%%%%%%%%%%%%%%%%%%%%%%%%%%%%%%%%%

%\appendix
%\section{Appendix Title}

%%% CALL LIST OF REFERENCES (natbib STYLE) % \bibliography{jkas-sample}
\bibliography{ms}

% \begin{thebibliography}{}

% %%% PUT YOUR REFERENCES HERE %%%%%%%%%%%%%%%%%%%%%%%%%%%%%%%%%%%%%%%%

% \bibitem[Salpeter(1955)]{salpeter1955} Salpeter, E. E. 1955, The Luminosity Function and Stellar Evolution, ApJ, 121, 161

% %%% END LIST OF REFERENCES %%%%%%%%%%%%%%%%%%%%%%%%%%%%%%%%%%%%%%%%%%

% \end{thebibliography}

\end{document}